\documentclass{emulateapj}

\newcommand{\etal}{et~al.}
\newcommand{\eg}{e.g., }
\newcommand{\ie}{i.e., }

\newcommand{\Msun}{M_{\odot}}

\newcommand{\ergs}{ergs~s$^{-1}$}

\newcommand{\Ed}{\dot{E}_{\rm dep}}
\newcommand{\Edep}{\dot{E}_{\rm dep,51}}

\newcommand{\Mms}{M_{\rm ms}}
\def\gsim{\mathrel{\rlap{\lower 4pt \hbox{\hskip 1pt $\sim$}}\raise 1pt
\hbox {$>$}}}
\def\lsim{\mathrel{\rlap{\lower 4pt \hbox{\hskip 1pt $\sim$}}\raise 1pt
\hbox {$<$}}}

\newcommand{\Mjet}{M_{\rm jet}}
\newcommand{\Mbh}{M_{\rm rem}}
\newcommand{\Et}{{E}_{\rm dep}}

\newcommand{\thj}{\theta_{\rm jet}}
\newcommand{\fth}{f_{\rm th}}
\newcommand{\Gj}{\Gamma_{\rm jet}}

\newcommand{\Min}{M_0}
\newcommand{\Rin}{R_0}
\newcommand{\Mfe}{M{\rm (Fe)}}

\begin{document}

\title{Aspherical Properties of Hydrodynamics and Nucleosynthesis in Jet-induced Supernovae}

\author{
 Nozomu~Tominaga\altaffilmark{1,2}
 }

\altaffiltext{1}{Optical and Infrared Astronomy Division, National
Astronomical Observatory, 2-21-1 Osawa, Mitaka, Tokyo 181-8588, Japan;
nozomu.tominaga@nao.ac.jp}
\altaffiltext{2}{Department of Astronomy, School of Science,
University of Tokyo, 7-3-1 Hongo, Bunkyo-ku, Tokyo, Japan}

\setcounter{footnote}{2}

\begin{abstract}
 Jet-induced supernovae (SNe) have been suggested to occur in gamma-ray
 bursts (GRBs) and highly-energetic SNe (hypernovae).
 I investigate hydrodynamical and nucleosynthetic properties of the
 jet-induced explosion of a population III $40\Msun$ star with a
 two-dimensional special relativistic hydrodynamical code. The abundance
 distribution after the explosion and the angular dependence of the
 yield are obtained for the models with high and low energy deposition rates
 $\Ed=120\times10^{51}$~\ergs\ and $1.5\times10^{51}$~\ergs. 
 The ejection of Fe-peak products and the fallback of unprocessed
 materials in the jet-induced SNe account for the abundance patterns of
 the extremely metal-poor (EMP) stars.
 It is also found that the peculiar abundance pattern of a
 Si-deficient metal-poor star HE~1424--0241 is reproduced by the
 angle-delimited yield for $\theta=30^\circ-35^\circ$ of the model with
 $\Ed=120\times10^{51}$~\ergs.  Furthermore, I compare the yield of
 the jet-induced explosion with that of the spherical explosion and
 confirm the ejection and fallback in the jet-induced explosion 
 is almost equivalent to the ``mixing-fallback''
 in spherical explosions. In contrast to the spherical models, however,
 the high-entropy environment is realized in the
 jet-induced explosion and thus [(Sc, Ti, V, Cr, Co, Zn)/Fe] are enhanced. 
 The enhancements of [Sc/Fe] and [Ti/Fe] improve agreements
 with the abundance patterns of the EMP stars.
\end{abstract}


\section{Introduction}
\label{sec:2Drelintroduction}

Gamma-ray bursts (GRBs) are phenomena emitting $\gamma$-ray for short
periods followed by a power-law decaying afterglow. The origin had been
covered for a long while, but it has become clear that
long GRBs are associated with supernovae (SNe). Three
GRB-associated SNe have been observed so far: GRB~980425/SN~1998bw
\citep{gal98}, GRB~030329/SN~2003dh \citep{sta03,hjo03}, and
GRB~031203/SN~2003lw \citep{mal04}. They are all energetic explosions of
massive stars, called hypernovae (\eg
\citealt{nom06,nom07} for a review).

Although the explosion mechanism of GRBs and GRB-associated SNe is still
uncovered, the following photometric and spectroscopic observations
indicate that they are aspherical explosions with jet(s). (1) The
light curve of the GRB afterglow has shown a
polychromatic break in the power-law decay. The break is called a ``jet
break'' explained by a deceleration of a relativistic jet and
relativistic beaming of light (\eg \citealt{fra01,pir05}). (2) The nebular
spectra of SN~1998bw show narrower [\ion{O}{1}] lines than
[\ion{Fe}{2}] lines \citep{pat01} indicating that O locates in the
inner and lower-velocity region than Fe. This is realized in an aspherical explosion but not in a
spherical explosion \citep{mae02,mae06a}.

The aspherical explosions are indirectly suggested from the abundance patterns
of extremely metal-poor (EMP) stars with [Fe/H] $<-3$.\footnote{Here [A/B] 
$=\log_{10}(N_{\rm A}/N_{\rm B})-\log_{10}(N_{\rm A}/N_{\rm B})_\odot$, 
where the subscript $\odot$ refers to the solar value and $N_{\rm A}$
and $N_{\rm B}$ are the abundances of elements A and B,
respectively.} Such EMP stars are likely to show the
nucleosynthesis yields of a single or a few core-collapse SN/SNe
\citep{aud95,bee05}. 
The abundance patterns are reproduced by the
``mixing-fallback'' models that assume the extensive mixing of the
shocked material before the fallback in a
spherical SN model \citep{ume02,ume03,ume05,iwa05,tom07b}. 
In particular, the abundance patterns of EMP stars with high
[Co/Fe] and [Zn/Fe] ($\sim0.5$) are explained by energetic explosion
models because the high ratios require explosive nucleosynthesis under
high entropy (\eg \citealt{ume05,tom07b}). Since the fallback
doesnot take place in energetic spherical or quasi-spherical explosions
(\eg \citealt{woo95,iwa05}),
the mixing and fallback are 
interpreted as a consequence of the aspherical explosion. Indeed,
the abundance patterns of the EMP stars are reproduced by the
jet-induced explosion model (\citealt{tom07a}). 

Since the jet-induced explosion contributes to many astronomical
phenomena, it is important to make a quantitative prediction on the
nucleosynthesis outcome of the jet-induced explosions. In order to
follow the jet propagation, fallback, and nucleosynthesis 
on the site and
compare the jet-induced explosion models with the EMP stars,
GRBs, and SNe, it is required to calculate multi-dimensional
hydrodynamics and nucleosynthesis with the gravity and the special
relativity.\footnote{In order to follow the core
collapse, jet formation, and nucleosynthesis on the site, it is also
required to involve the neutrino, the general relativity, and possibly
the magnetic field (\eg \citealt{pru05,fro06,jan07}).} 
Although there are many studies on
the GRB jet propagation of the stellar mantle using the special
relativistic hydrodynamics (\eg \citealt{alo00,zha04,miz06}),
they did not include the gravity or calculate nucleosynthesis.

To investigate the yields of the jet-induced SNe, it is
crucial to include the gravity because the fallback plays an important
role on the nucleosynthesis yields. The studies by the use of Newtonian
calculations have concluded
that the energy deposition rate ($\Ed$) sensitively affects SN
nucleosynthesis \citep{mae03b,nag06}. This result has been confirmed for
the special relativistic cases \citep{tom07a}. In particular,
\cite{tom07a} have shown that the jet-induced explosions
with various $\Ed$ can explain both the variations of 
GRB-associated SNe and the EMP, carbon-enhanced EMP (CEMP, [C/Fe]~$>1$),
and hyper metal-poor (HMP, [Fe/H]~$<-5$) stars. 

The previous studies have proved mostly the angle-integrated yields and
shown that the abundance patterns of the EMP stars are reproduced by the
angle-integrated yield. However, the abundance distribution of the
jet-induced explosion depends on the direction (\eg \citealt{mae03b}). 
Thus, the abundance patterns of the next-generation stars might
depend on the direction. 
I calculate aspherical stellar explosions 
induced by highly relativistic jets and obtain hydrodynamical and
nucleosynthetic structures of such jet-induced explosion models. In
particular, I investigate the angular dependence of the yield to compare
the yields with the abundance patterns of the metal-poor stars.
Furthermore, I compare the jet-induced explosion with the spherical
SN model applied the mixing-fallback model and connect properties of the
jet-induced explosion to the mixing-fallback model.

In \S~\ref{sec:model}, the applied models are described. In
\S~\ref{sec:hyd}, I present the hydrodynamical and nucleosynthetic
structures of the jet-induced explosion model, investigate the angular dependence
of the yields, and compared the jet-induced explosion model with the spherical
SN model. In \S~\ref{sec:summary}, the
conclusion and discussion are presented. In Appendixes, the
hydrodynamics and nucleosynthesis code is described and tested.

\section{Models}
\label{sec:model}

I investigate a jet-induced SN explosion of a Pop III $40\Msun$ star
\citep{ume05,tom07b} by means of a two-dimensional relativistic Eulerian
hydrodynamic and nucleosynthesis calculation with the gravity
(Appendix~\ref{sec:hydtest}). The nucleosynthesis calculation is
performed as a post-processing \citep{hix96,hix99} with the reaction
network including 280 isotopes up to $^{79}$Br (see Table~1 in
\citealt{ume05}). The thermodynamic histories are traced by maker
particles representing Lagrangian mass elements (\eg
\citealt{hac90,mae03b}, see also Appendix~\ref{sec:nuctest}). 
A computational domain initially ranges up to the stellar surface where
$R_{\rm star}=2\times10^7~{\rm km}$ and is captured by 200 logarithmical
grids in the $r$-direction and 100 uniform grids in the
$\theta$-direction. A circumstellar matter (CSM) extends from the
stellar surface with the slope $\rho\propto r^{-2}$. 

\begin{figure}
\epsscale{.9}
\plotone{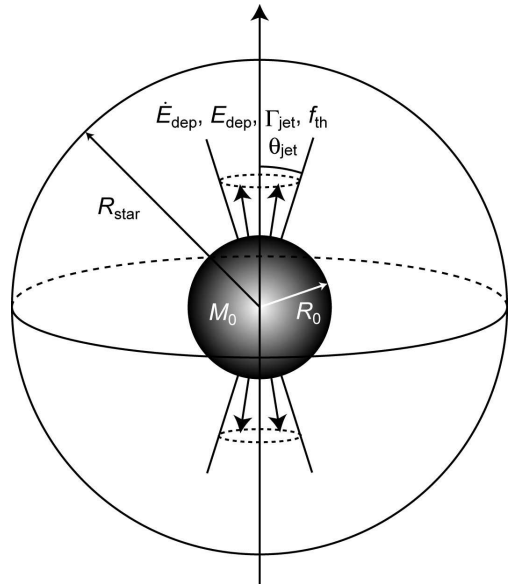}
\caption{Schematic picture of the jet-induced explosion. \label{fig:2DrelJet}}
\end{figure}

The explosion mechanism of GRB-associated SNe is still under debate (\eg a neutrino
annihilation, \citealt{woo93a,mac99}; and a magneto-rotation,
\citealt{bla77,bro00,mizn04}). Thus, I do not consider how the jet is
launched, but the jet is dealt parametrically with the following five
parameters (Fig.~\ref{fig:2DrelJet}): energy deposition rate ($\Ed$),
total deposited energy ($\Et$), initial half angle of the jets ($\thj$),
initial Lorentz factor ($\Gj$), and the ratio of thermal to total
deposited energies ($\fth$).
The jet is injected from the inner boundary at an enclosed mass $\Min$
corresponding to a radius $\Rin$. The density, pressure, and
velocities of the jet are described with the five parameters
(Appendix~\ref{sec:jetin}) and put as a 
boundary condition at the inner boundary.
The jet is assumed to consist of the
accreted matter and to expand adiabatically below the inner boundary.
After the jet is injected into the computational domain, the
thermodynamic histories are traced by the marker particles. 

\begin{deluxetable*}{c|ccc|cccc|c}
 \tabletypesize{\scriptsize}
 \tablecaption{Jet-induced explosion models. \label{tab:2Drelmodel}}
 \tablewidth{0pt}
 \tablehead{
   \colhead{Name}
 & \colhead{$\Min$}
 & \colhead{$\Rin$}
 & \colhead{$\Ed$}
 & \colhead{$\Et$}
 & \colhead{$\thj$}
 & \colhead{$\Gj$}
 & \colhead{$\fth$}
 & \colhead{$\Mbh$}\\
   \colhead{}
 & \colhead{[$\Msun$]}
 & \colhead{[km]}
 & \colhead{[$10^{51} {\rm ergs~s^{-1}}$]}
 & \colhead{[$10^{51} {\rm ergs}$]}
 & \colhead{[degrees]}
 & \colhead{}
 & \colhead{}
 & \colhead{[$\Msun$]}
 }
\startdata
 A& 1.4 & 900  & 120& 15 & 15 & 100 & 10$^{-3}$ & 9.1 \\
 B& 1.4 & 900  & 1.5& 15 & 15 & 100 & 10$^{-3}$ &16.9 \\
 C& 2.3 & 2700 & 120& 15 & 15 & 100 & 10$^{-3}$ & 8.1 
\enddata
\end{deluxetable*}

In this paper, I show three models; (A) a model with $\Edep=\Ed/(10^{51}{\rm ergs~s^{-1}})=120$ and
$\Min=1.4\Msun$ ($\Rin=900$ km), (B) a model with $\Edep=1.5$ and
$\Min=1.4\Msun$ ($\Rin=900$ km), and
(C) a model with $\Edep=120$ and $\Min=2.3\Msun$ ($\Rin=2700$ km). The other parameters
are same for each model; $\Et=1.5\times10^{52}$
ergs,\footnote{\cite{fra01} suggested the $\gamma$-ray energies
($E_\gamma$) of GRBs are clustered at $E_\gamma\sim5\times10^{50}$ ergs. 
Although $E_\gamma$ is 30 times smaller than $\Et$, there are large
uncertainties on the relation between $E_\gamma$ and $\Et$. For example, 
the energy possessed by the relativistic
matter depends on the interactions between the relativistic jet and the
stellar mantle and is reduced compared to the injected energy
\citep{zha03,zha04}. Moreover, the radiative conversion efficiency of
a kinetic energy depends on an unknown $\gamma$-ray emission
mechanism. The efficiency is estimated to be $\sim2-40\%$ for collisions of
internal shocks \citep{kob97} or $>50\%$ for nonthermalized photospheric
emissions \citep{iok07}. $\Et$ adopted in this paper might be appropriate for
the former emission mechanism or excessive for the latter emission
mechanism. The dependence of the jet-induced explosions on $\Et$ 
will be studied in future.} $\thj=15^\circ$,
$\Gj=100$ and $\fth=10^{-3}$. The mass of jets is $\Mjet\sim8\times10^{-5}\Msun$.
The model parameters and the central remnant mass ($\Mbh$) 
are summarized in Table~\ref{tab:2Drelmodel}.
Models A and B are used in \cite{tom07a} and they reproduce
the abundance patterns of the EMP stars\footnote{An averaged abundance
pattern of four EMP stars, CS~22189--009, CS~22172--002,
CS~22885--096, and CD~$-$38~245, is adopted.} \citep{cay04} and HE~1327--2326 (\eg
\citealt{fre05,fre06,col06}), respectively.

\section{Results}
\label{sec:hyd}

\begin{figure}
\epsscale{1.05}
  \plotone{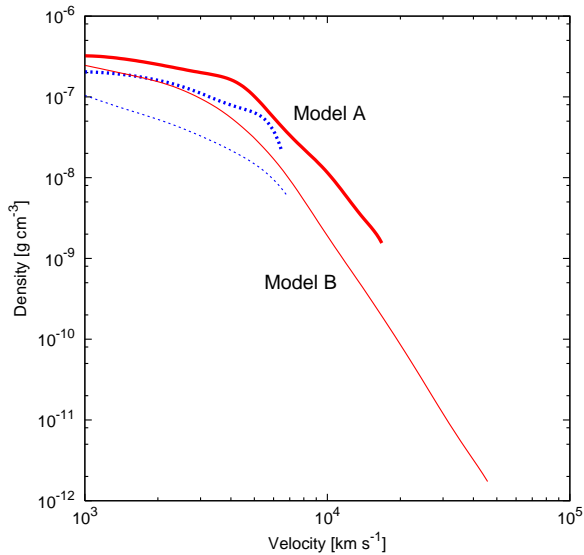}
 \caption{Density structures at $t=10^5$~s along the jet axis ({\it
 solid line}) and the equatorial plane ({\it dashed line}) of model A
 ({\it thick line}) and B ({\it thin line}). \label{fig:2Drelvrho}}
\end{figure}

The hydrodynamical calculations are followed until the homologously
expanding structure is reached ($v\propto r$). Then, the ejected mass
elements are identified from whether their radial velocities exceed the
escape velocities at their positions. The density structures of models A
and B at $t=10^5$~s are shown in Figure~\ref{fig:2Drelvrho}. 
The density along the jet axis is higher than the density along the equatorial
plane because the matter is more easily ejected along the jet axis
than along the equatorial plane. And the SN
ejecta of model A is denser and more compact than that of model B
because the ejected mass of model A is larger than that of model B.

\subsection{Fallback}
\label{sec:fallback}

\begin{figure}
\epsscale{1.}
  \plotone{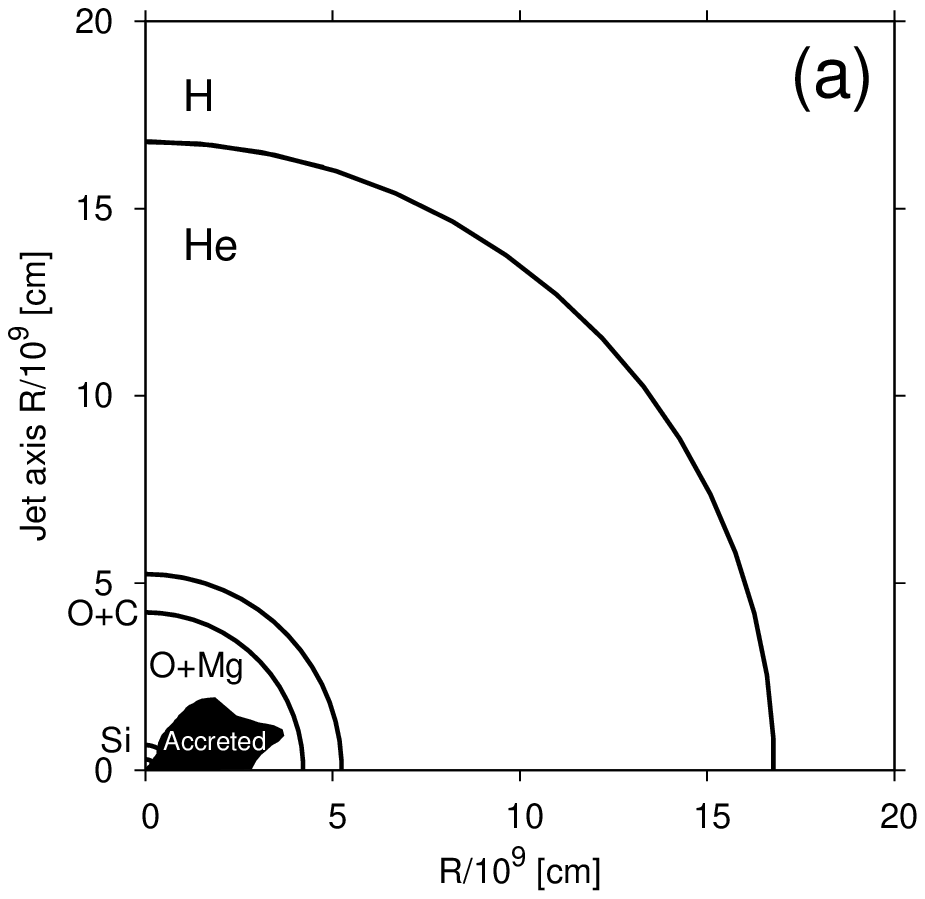}
  \plotone{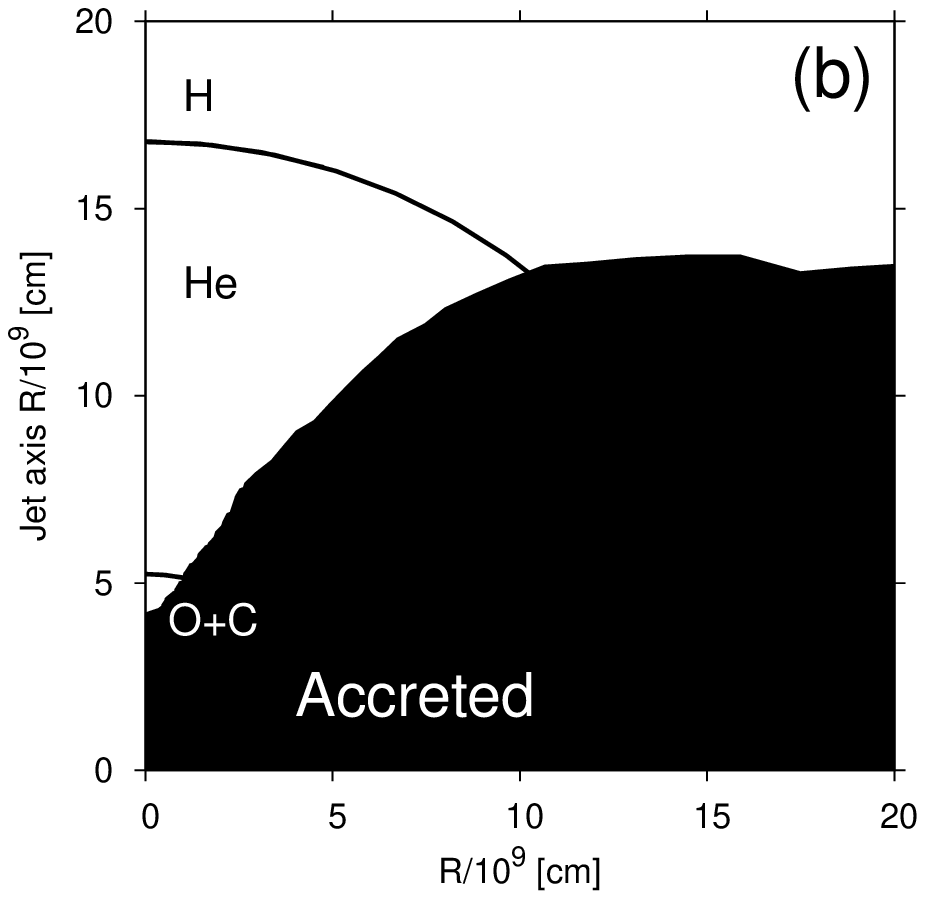}
 \caption{Initial locations of the mass elements which are finally
 accreted ({\it black}), for (a) model A and
 (b) model B. The background circles represent the boundaries between the layers in the
 progenitor star; the H, He, O+C, O+Mg, and Si layers from the outside. \label{fig:2Drelfallback}}
\end{figure}

Figures~\ref{fig:2Drelfallback}a and \ref{fig:2Drelfallback}b show
``accreted'' regions for models A and B, where the accreted mass
elements initially located in the progenitor. The O layer is
separated into the two layers: (1) the O+Mg layer with 
$X({\rm ^{24}Mg})>0.01$ and (2) the O+C layer with $X({\rm ^{12}C})>0.1$. 
The inner matter is ejected along the jet-axis but not along the
equatorial plane. On the other hand, the outer matter is ejected even
along the equatorial plane, since the lateral expansion of the shock
terminates the infall as the shock reaches the equatorial plane.

The accreted mass is larger for lower $\Ed$. This stems from the balance
between the ram pressures of the injecting jet ($P_{\rm jet}$) and the
infalling matter ($P_{\rm fall}$) (\eg \citealt{fry99,fry03,mae06c}). 
While $P_{\rm jet}$ depends on $\Ed$, $P_{\rm fall}$ depends on the
presupernova structure. If the matter falls freely, $P_{\rm fall}$ is
proportional to $\rho_r r^{3/2}M_r$ for the matter having located at $r$ 
and $M_r$ in the progenitor where the presupernova density is $\rho_r$.
The critical energy deposition rate ($\dot{E}_{\rm dep,cri}$) giving  
$P_{\rm jet}(\dot{E}_{\rm dep,cri})=P_{\rm fall}(M_r)$ is lower 
for the outer layer (\ie larger $M_r$, \citealt{fry03,mae06c}).
Therefore, the jet injection with lower $\Ed$ is enabled at a later time
when the central remnant becomes more massive. Additionally, the lateral
expansion of the jet is more efficiently suppressed for lower $\Ed$. As
a result, the accreted region and $\Mbh$ are larger for lower $\Ed$. 

A model with lower $\Ed$ has larger $\Mbh$, higher [C/O], [C/Mg], and
[C/Fe], and smaller $\Mfe$ because of the larger amount of fallback \citep{tom07a}. 
The larger amount of fallback decreases the mass of the inner core
(\eg Fe, Mg, and O) relative to the mass of the outer layer (\eg C,
Figs.~\ref{fig:2Drelfallback}ab). Since O and Mg are synthesized in the
inner layers than C, [C/O] and [C/Mg] are larger for the
larger infall of the O layer. Also, the fallback of the O layer decreases
$\Mfe$ because Fe is mainly synthesized explosively in the Si and O+Mg layers.
Therefore, the variation of $\Ed$ in the jet-induced explosions predicts that
the variations of [C/O], [C/Mg], and [C/Fe] are corresponding to
the variation of $\Mfe$.

\subsection{Abundance distribution}
\label{sec:angledep}

\begin{figure}
\epsscale{1.2}
 \plotone{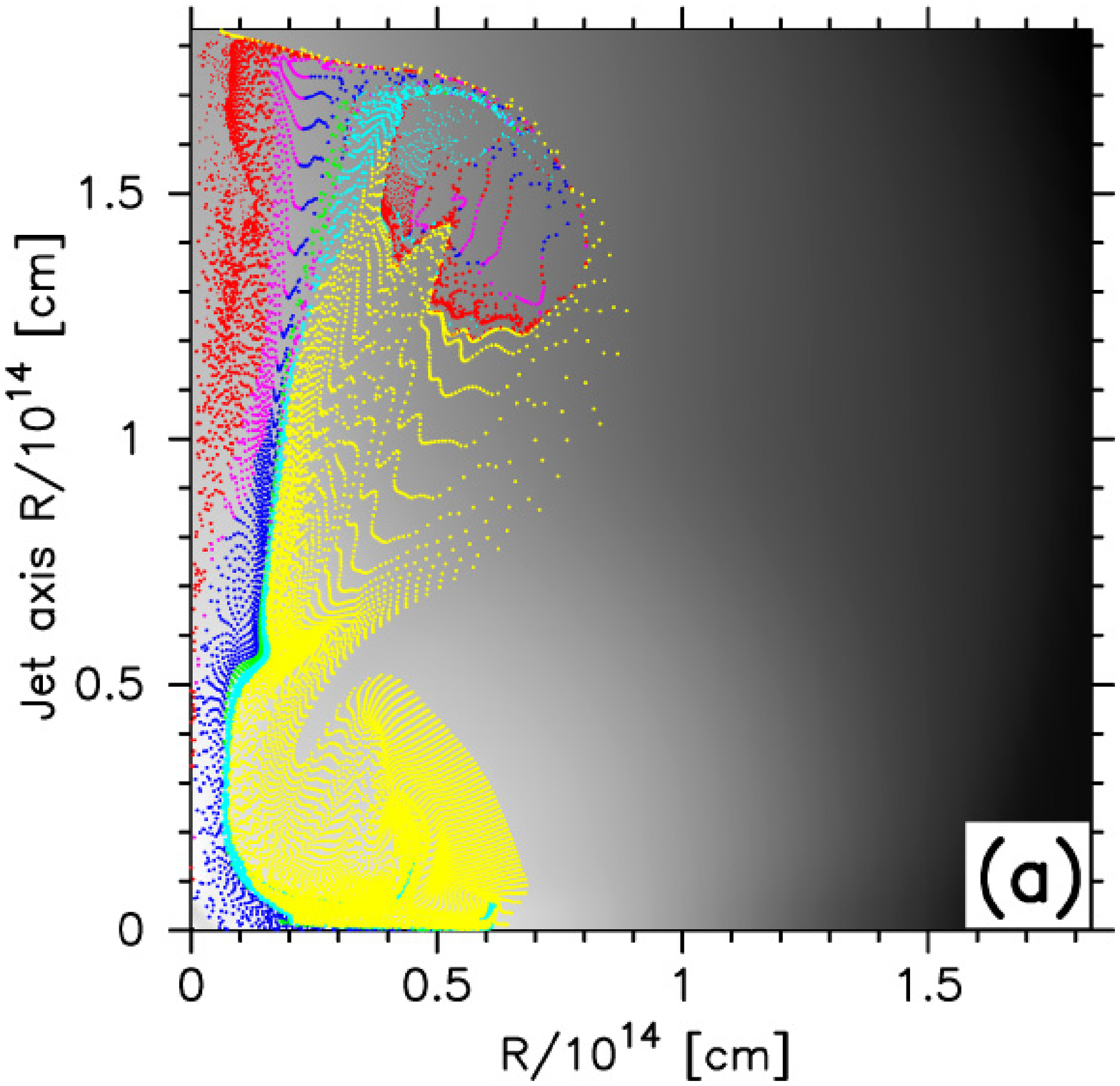}
 \plotone{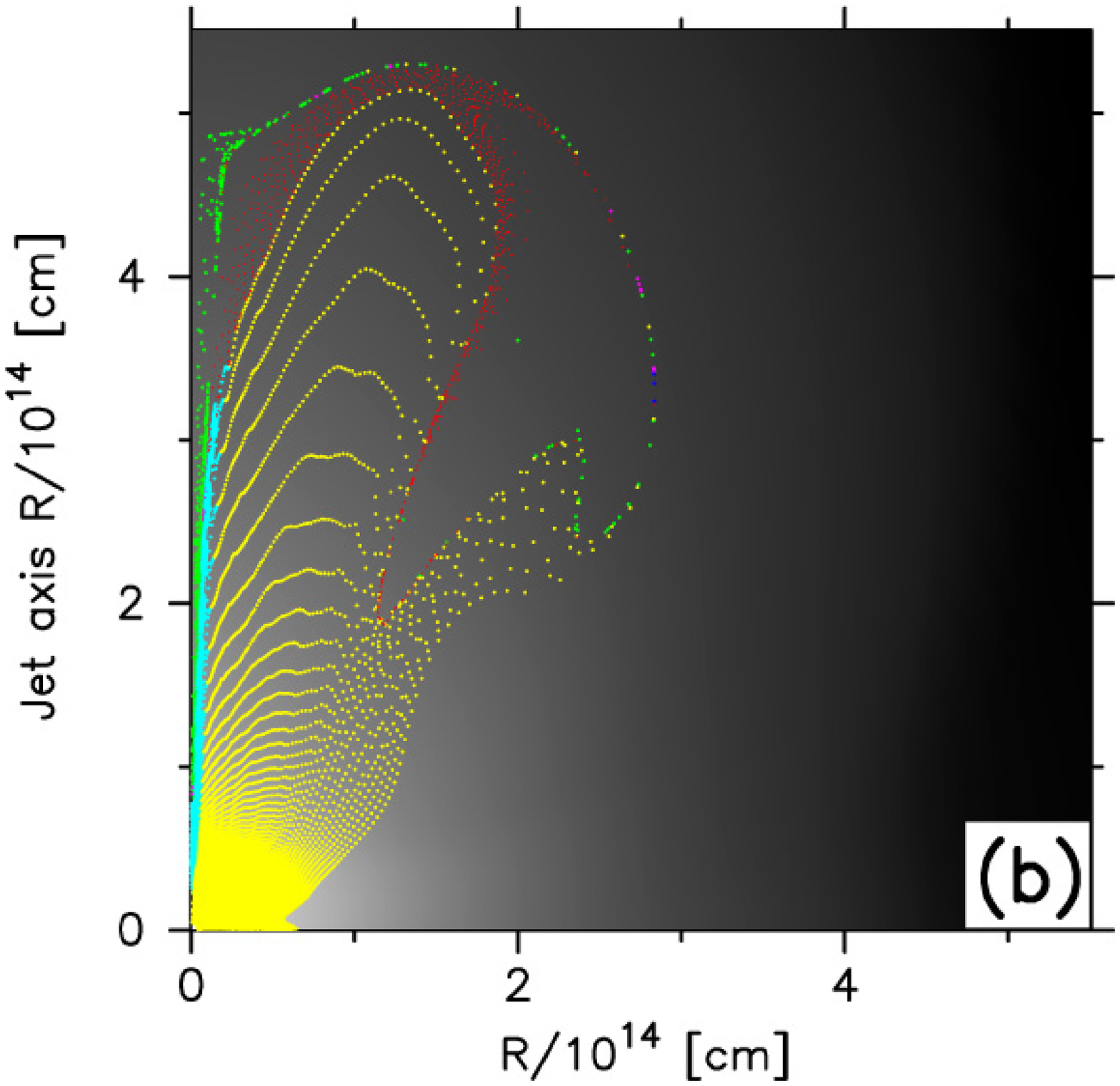}
 \caption{Density structure ({\it back ground
 gray scale}) and the positions of the mass elements at $t=10^5$~s
 for (a) model A and (b) model B. Color of the marks
 represents the abundance of the mass element (H: {\it yellow}, He: {\it
 cyan}, O+C: {\it green}, O+Mg: {\it blue}, Si: {\it magenta}, and Fe: {\it
 red}). Size of the marks represents the origin of the mass element
 (the jet: {\it dots}, and the shocked stellar mantle: {\it filled circles}). \label{fig:2Drelej}}
\end{figure}

Figures~\ref{fig:2Drelej}a and \ref{fig:2Drelej}b show the abundance distributions and
density structures at $t=10^5$~s for models A and
B. I classify the mass elements by their abundances as follows:
(1) Fe with $X({\rm ^{56}Ni})>0.04$, (2) Si with $X({\rm ^{28}Si})>0.08$, 
(3) O+Mg with $X({\rm ^{16}O})>0.6$ and $X({\rm ^{24}Mg})>0.01$, 
(4) O+C with $X({\rm ^{16}O})>0.6$ and $X({\rm ^{12}C})>0.1$, 
(5) He with $X({\rm ^{4}He})>0.7$, and (6) H with $X({\rm ^{1}H})>0.3$.
If a mass element satisfies two or more conditions, the mass element
is classified into the class with the smallest number.

The abundance distribution and thus the composition of the ejecta depend
on the direction. In model A, the O+Mg, O+C, He, and H mass elements
locate in the all direction. On the other hand, most of the Fe and Si
mass elements locate at $\theta<10^\circ$ and stratify in this order
from the jet axis and a part of them locate at
$15^\circ<\theta<35^\circ$. Interestingly, the Fe mass elements surround
the Si mass elements at $15^\circ<\theta<35^\circ$. In model B, most of
the O+C and He mass elements locate at $\theta<3^\circ$ because they
are ejected only along the jet axis, while the Fe
mass elements injected as a jet expand laterally up to
$\theta\sim50^\circ$ and the H mass elements are distributed in the all
directions. The lateral spread of the Fe mass elements in
models A and B are led by the collision with the stellar mantle and
the internal pressure of the jet.

\subsubsection{Angular dependence of the yields}

\begin{figure}
\epsscale{1.1}
  \plotone{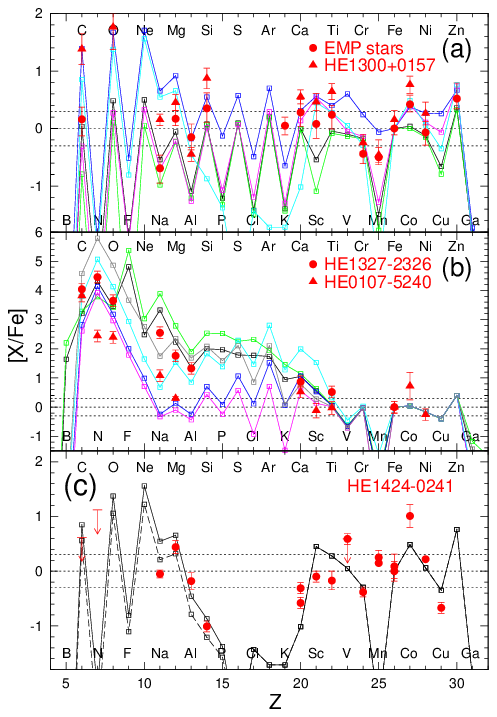}
 \caption{Comparisons between the abundance patterns of the
 angle-integrated yield ({\it black}), the EMP, CEMP, and HMP stars 
 ({\it red filled marks},
  \citealt{cay04,fre07,chr02,bes05,col06,fre05,fre06}), and the
 angle-delimited yields ({\it colored lines})
 for (a) model A and (b) model B. The color of the lines
 represent the yields integrated over $0^\circ\leq\theta<10^\circ$ ({\it
  green line}),
 $10^\circ\leq\theta<20^\circ$ ({\it blue line}),
 $20^\circ\leq\theta<30^\circ$ ({\it magenta line}),
 $30^\circ\leq\theta<40^\circ$ ({\it cyan line}), and
 $40^\circ\leq\theta<50^\circ$ ({\it gray line}). (c) Comparison between the
 abundance pattern of HE~1424--0241 ({\it red filled circles}, \citealt{coh06}) and the
 angle-delimited yields of model A for $30^\circ\leq\theta<40^\circ$ ({\it
 solid line}) and $30^\circ\leq\theta<35^\circ$ ({\it dashed line}). \label{fig:2Drelangledep}}
\end{figure}

I investigate the angle-delimited yields integrated over each $10^\circ$
for models A and B, although the integration range might be too wide to
be taken in a single next-generation star.
Fe is mainly distributed at $\theta<40^\circ$ for model A (Fig.~\ref{fig:2Drelej}a)
and $\theta<50^\circ$ for model B (Fig.~\ref{fig:2Drelej}b).
Figures~\ref{fig:2Drelangledep}a and \ref{fig:2Drelangledep}b
show the abundance patterns of the angle-delimited yields for
$\theta<40^\circ$ of model A and $\theta<50^\circ$ of model B, respectively.
The yields are compared with the abundance patterns of the EMP
stars \citep{cay04}, the CEMP stars (HE~1300+0157, \citealt{fre07}),
 and the HMP stars (HE~0107--5240, \eg
\citealt{chr02,bes05,col06} and HE~1327--2326, \eg \citealt{fre05,fre06,col06}).

Figure~\ref{fig:2Drelangledep}a shows the angle-delimited yields of
model A. The abundance patterns of the angle-delimited yields are
determined by which mass elements are included into the integration.
Because of the stratified abundance distribution, the yields for
$0^\circ\leq\theta<10^\circ$ and $10^\circ\leq\theta<20^\circ$ show low
[C/Fe] and [O/Fe] and high [C/Fe] and [O/Fe], respectively.
Intriguingly, the region with $30^\circ\leq\theta<40^\circ$ includes the Fe mass
elements in the outer layer and the O+Mg and O+C mass elements in the
inner layer but not the Si mass elements. Thus, the yield for
$30^\circ\leq\theta<40^\circ$ shows a Si-deficient abundance pattern.
Furthermore, the Fe and Si mass elements at $15^\circ<\theta<35^\circ$
initially located along the jet axis ($\theta<10^\circ$), and thus the
high-entropy environment is realized in these mass
elements and [Sc/Fe], [Ti/Fe], [Co/Fe], and [Zn/Fe] are enhanced in
the yields for $10^\circ\leq\theta<40^\circ$. 
On the other hand, [Sc/Fe] is low in the yield for
$0^\circ\leq\theta<10^\circ$ because the matter initially located at
$10^\circ\leq\theta$.

Figure~\ref{fig:2Drelangledep}b shows the angle-delimited yields of
model B. The all angle-delimited yields of model B show high [C/Fe]. 
This is the same feature as the angle-integrated yield. Since most of the
heavy elements locate at $\theta<10^\circ$, the abundance pattern of the
yield for $0^\circ\leq\theta<10^\circ$ is similar to that of the
angle-integrated yield, except for N that is mostly contained in the H
mass elements. On the other hand, the yields for
$10^\circ\leq\theta<50^\circ$ consist of the Fe mass elements injected
as a jet and the O+C mass elements. In the yields for
$10^\circ\leq\theta<50^\circ$, [C/Mg] is almost similar and the
differences of [C/Fe] and [Mg/Fe] mainly stem from the different amount
of Fe. 

\subsubsection{Abundance patterns of the metal-poor stars}

A very peculiar, Si-deficient, metal-poor star HE~1424--0241
was observed \citep{coh06}. Its abundance pattern with high [Mg/Si] ($\sim1.4$)
and normal [Mg/Fe] ($\sim0.4$) is difficult to be reproduced by
previous SN models. This is because 
$\log\{[X({\rm Mg})/X({\rm Si})]/[X({\rm Mg})/X({\rm Si})]_\odot\}\lsim1.6$ 
is realized in the O+Mg layer at the presupernova stage (\eg
\citealt{woo95,ume05}). Thus, in order to reproduce
the abundance pattern of HE~1424--0241, the SN yield is required to
include explosively-synthesized Fe but not explosively-synthesized Si.

The angle-delimited yield may possibly explain the high
[Mg/Si] and normal [Mg/Fe] (Fig.~\ref{fig:2Drelangledep}c). Figure~\ref{fig:2Drelangledep}c
shows that the yields integrated over $30^\circ\leq\theta<40^\circ$ and
$30^\circ\leq\theta<35^\circ$ of model A reproduce the abundance pattern of HE~1424--0241. 
The yield consist of Mg in the inner region and Fe in the outer region
(Fig.~\ref{fig:2Drelej}a).
Although there are some elements to be improved, the elusive feature of
HE~1424--0241 could be explained by taking into account the angular
dependence of the yield.  The high [Mg/Si] and normal [Mg/Fe] can be
realized with an appropriate integration range if the Fe mass elements
penetrate the stellar mantle (\ie the duration of the jet injection is
long) and if the O+Mg mass elements are ejected in all directions (\ie
$\Ed$ is high).

If the yield depends on the direction, the abundance patterns of the
angle-delimited yields have a large scatter
(Fig.~\ref{fig:2Drelangledep}a). 
Although the diversities of [C/Fe] and [O/Fe] in the angle-delimited
yields are similar to those in the observations of the EMP and CEMP
stars, the observations of the EMP stars provide abundance
ratios with comparatively small scatters (\eg [Sc/Fe], \citealt{cay04}). This
implies that SNe contributing to most of the EMP stars experienced strong
mixing of the ejecta, \eg due to an interaction with interstellar
medium (ISM), so that the yield no longer depends on the direction. 
Actually, it was suggested that the interaction between the SN ejecta
and ISM induces further mixing of the abundances if ISM has an
inhomogeneous structure (\eg \citealt{nak00}). According to
structure formation calculations (\eg \citealt{yos07}), the Pop
III stars are surrounded by the inhomogeneous ISM. Therefore, the mixing of
the SN ejecta is likely to take place. On the other hand, the
abundance pattern of HE~1424--0241 implies weak mixing of the SN
ejecta. Such a different star formation process might cause the peculiar
abundance pattern of HE~1424--0241. In
order to conclude the origin of HE~1424--0241, it is necessary to
calculate three-dimensional evolution of a supernova remnant in the
inhomogeneous ISM. 

\subsection{Comparison with the spherical supernova model}
\label{sec:MFjet}

\begin{figure*}
\epsscale{.5}
  \plotone{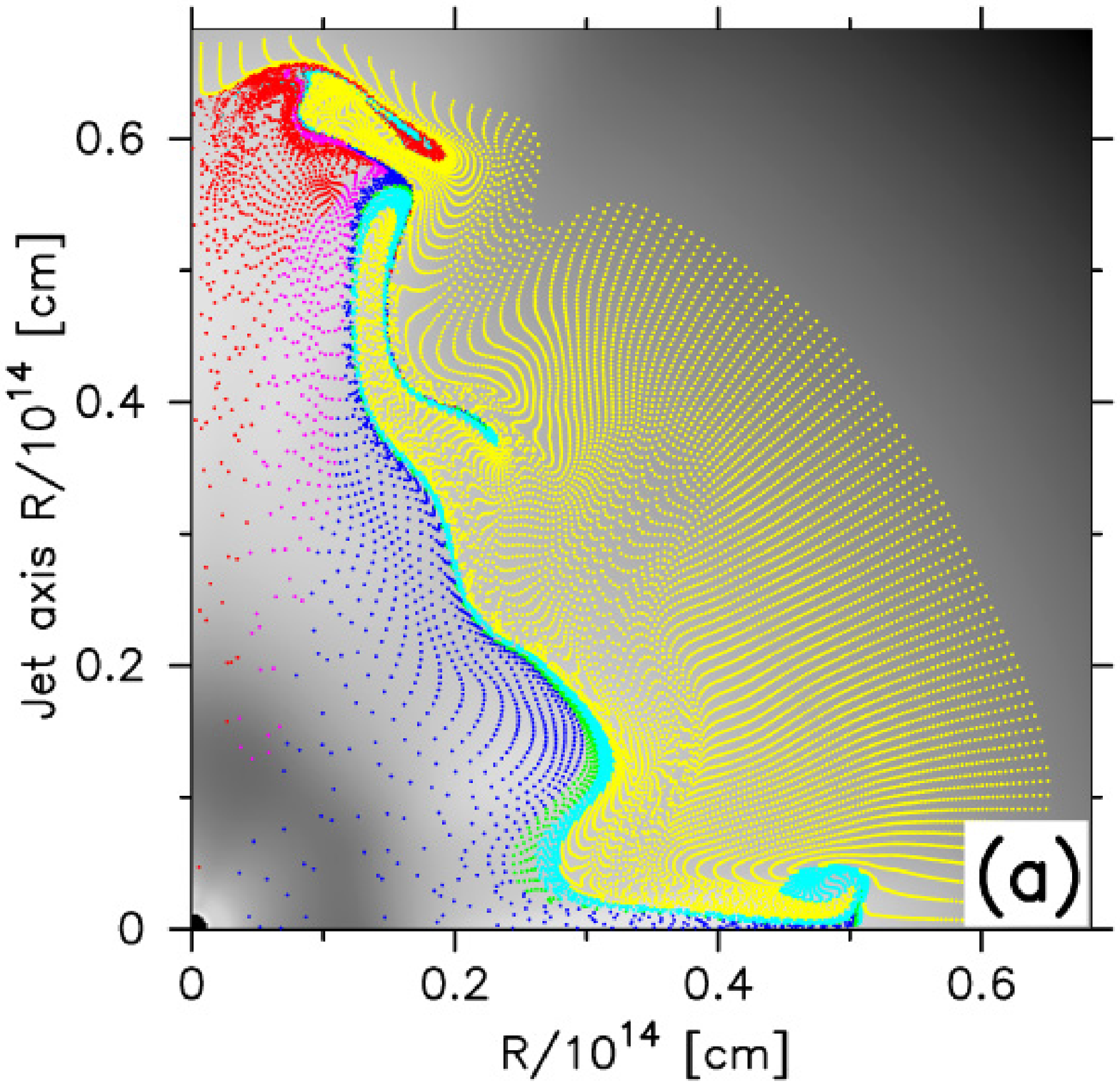}
\epsscale{.45}
  \plotone{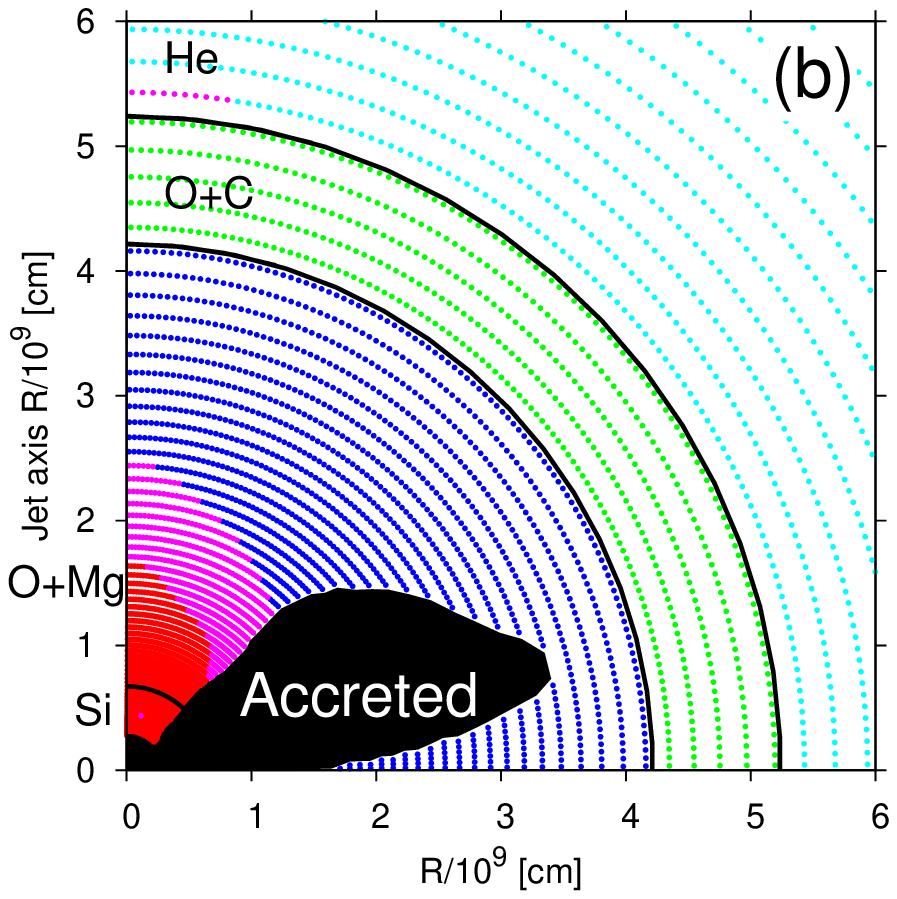}
\epsscale{.75}
  \plotone{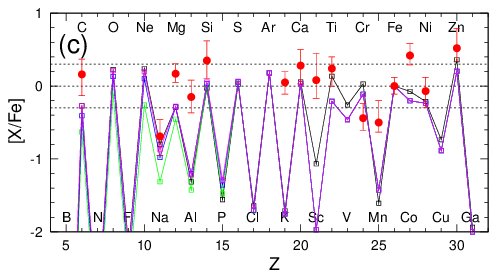}
 \caption{(a) Density structure ({\it back ground
 gray scale}) and the positions of the mass elements at $t=10^5$~s
 for model C. The color and size of the marks represent the same as
 Figs.~\ref{fig:2Drelej}ab.
 (b) Initial locations of the mass elements which are finally
 accreted ({\it black}) and ejected ({\it colored points}) for model
 C. The color of the mass elements represents the same as
 Figs.~\ref{fig:2Drelej}ab. (c) Comparison of the abundance patterns of
 model C ({\it black line}), the EMP stars ({\it red circles},
  \citealt{cay04}), and the mixing-fallback models with 
(1) $M_{\rm cut}{\rm (ini)}=2.3\Msun$, 
$M_{\rm mix}{\rm (out)}=12.2\Msun$ and $f=0.41$ ({\it green line}), (2) $M_{\rm cut}{\rm (ini)}=2.3\Msun$, 
$M_{\rm mix}{\rm (out)}=10.3\Msun$ and $f=0.27$ ({\it blue line}), and (3) $M_{\rm cut}{\rm (ini)}=2.3\Msun$, 
$M_{\rm mix}{\rm (out)}=10.8\Msun$ and $f=0.19$ ({\it magenta line}). \label{fig:2DrelEdotfallP3}}
\end{figure*}

The calculations of the jet-induced explosions show that the ejection
of the inner matter is compatible with the fallback of the outer matter
(Figs.~\ref{fig:2Drelfallback}ab). This is consistent with the
two-dimensional illustration of the mixing-fallback model (Fig.~12b in
\citealt{tom07b}). 

In this subsection, I clarify the relation between the mixing-fallback
model and the jet-induced explosion model by comparing the yields.
The mixing-fallback model has three parameters; initial mass cut
[$M_{\rm cut}{\rm (ini)}$], outer boundary of the mixing region 
[$M_{\rm mix}{\rm (out)}$], and a fraction of matter ejected from the
mixing region ($f$). The remnant mass is written as 
\begin{equation}
 \Mbh=M_{\rm cut}{\rm (ini)}+(1-f)[M_{\rm mix}{\rm (out)} - M_{\rm cut}{\rm (ini)}].
\end{equation}
The three parameters would relate to the hydrodynamical properties of
the jet-induced explosion models, \eg the inner boundary ($\Min$),
the outer edge of the accreted region ($M_{\rm acc,out}$),
and the width between the edge of the accreted region and the jet axis.

I apply model C for the comparison.
The inner boundary, the outer edge of the
accreted region, and the central remnant mass of model C are
$\Min=2.3\Msun$, $M_{\rm acc,out}=12.2\Msun$, and $\Mbh=8.1\Msun$. The
abundance distribution of model C at $t=10^5$~s is shown in
Figure~\ref{fig:2DrelEdotfallP3}a.

Figure~\ref{fig:2DrelEdotfallP3}b shows the initial positions of the
ejected mass elements in model C. Color of the marks represents
the composition of the mass elements. Explosive
nucleosynthesis takes place in the mass elements between the edge of the
accreted region and the jet axis. Therefore, the amounts of the explosive
nucleosynthesis products are reduced relative to the
spherical explosions. The shock collision at the
equatorial plane realizes the mass ejection from the deep O+Mg layer
along the equatorial plane. Although the mass ejection could be due to
the reflective boundary condition on the equatorial plane, this does not
affect the following discussion
because the amount of the mass ejection is small.

The angle-integrated yield of model C is compared with
the yields of the spherical SN model with
$\Mms=40\Msun$ and $\Et=3\times10^{52}$ ergs. For the spherical SN model, the
explosion energy is deposited instantaneously as a thermal bomb. 
Here, I set $M_{\rm cut}{\rm (ini)}$
to be the same as $\Min$ [\ie $M_{\rm cut}{\rm (ini)}=2.3\Msun$] and apply
three sets of parameters, $M_{\rm mix}{\rm (out)}$ and $f$, as follows:
 
(1) $M_{\rm mix}{\rm (out)}=12.2\Msun$ and $f=0.41$. 
$M_{\rm mix}{\rm (out)}$ corresponds to
$M_{\rm acc,out}$ and $f$ is the ratio of the
ejected mass and the total mass between $\Min$ and $M_{\rm acc,out}$.

(2) $M_{\rm mix}{\rm (out)}=10.3\Msun$ and $f=0.27$. 
$f$ is the fraction of the solid angle of the region where the ejected mass
elements located at $r\sim10^3~{\rm km}$ (the Si-burning region,
hereafter the fraction is written as $f_{\rm Si}$).
$M_{\rm mix}{\rm (out)}$ is set to yield the same $\Mbh$ as model C.

(3) $M_{\rm mix}{\rm (out)}=10.8\Msun$ and $f=0.19$. 
$f$ and $M_{\rm mix}{\rm (out)}$ are set to yield the most resemblant
abundance pattern to model C. The resultant
$\Mbh$ ($=9.2\Msun$) is slightly larger than $\Mbh$ of
model C. 

Figure~\ref{fig:2DrelEdotfallP3}c shows that the angle-integrated abundance
pattern of model C is roughly reproduced by the mixing-fallback
model with the parameter sets of (2) and (3). Thus, I conclude that
the jet-induced explosion is approximated by the mixing-fallback model reasonably.
$\Min$, $f_{\rm Si}$, and $\Mbh$ in the jet-induced explosion model are represented by 
$M_{\rm cut}{\rm (ini)}$, $f$, and $\Mbh$ in the mixing-fallback model,
respectively.

There are some elements showing differences, Sc, Ti, V, Cr, Co, and
Zn. The enhancements of [Sc/Fe] and [Ti/Fe] improve agreements with the
observations. The differences stem from the high-entropy explosion due to the
concentration of the energy injection in the jet-induced explosion (\eg \citealt{mae03b}). 
In particular, Sc is the most sensitive element to the entropy and 
[Sc/Fe] is more enhanced for the deeper $\Min$ because of the weaker lateral
expansion (Fig.~\ref{fig:2Drelangledep}a). 
Such thermodynamical features of the jet-induced explosion model cannot
be reproduced by the mixing-fallback model exactly but a ``low-density''
modification might mimic the high-entropy environment (\eg
\citealt{ume05,tom07b}).

\section{Conclusions and Discussion}
\label{sec:summary}

I perform two-dimensional hydrodynamical and nucleosynthesis
calculations of the jet-induced explosions of a Pop III $40\Msun$ star.
Here I test three jet-induced explosion models A, B, and C as summarized in
Table~\ref{tab:2Drelmodel} and conclude as follows.

(1) {\bf Fallback}: The dynamics and the abundance distributions depend sensitively on
the energy deposition rate $\Ed$. The explosion with lower $\Ed$
leads to a larger amount of fallback, and consequently smaller $\Mfe$
and higher [C/O],
[C/Mg], and [C/Fe]. Such dependences of [C/Fe] and $\Mfe$ on $\Ed$
predict that higher [C/Fe] tends to be realized for
lower [Fe/H]. This is consistent with the observations (\eg
\citealt{bee05}). Note, however, the formation
of star with low [C/Fe] and [Fe/H] is possible because [Fe/H] depends on the
swept-up H mass, \ie the interaction between the SN ejecta and ISM (\eg 
\citealt{cio88}).

(2) {\bf Abundance distribution and angular dependence}: I present the
aspherical abundance distributions and investigate the angular
dependence of the yield. After the explosion, the elements ejected even
along the equatorial plane locate in the all directions, while the
elements ejected only along the jet axis expand laterally but tend to be
distributed along the jet axis. The abundance distributions in
the jet-induced SN ejecta could be examined by spatially-resolved
observations of supernova remnants (\eg \citealt{hwa04,keo07}).

The angle-delimited yield could reproduce the extremely peculiar
abundance pattern of HE~1424--0241. However, the angle-delimited yields
of model A have a large scatter that may be inconsistent with the
relatively small scatter in the abundance ratios of the EMP
stars. This implies that the angular dependence of the yield in most SNe
is diluted by the strong mixing of the SN ejecta. 
On the other hand, the angle-delimited yields of model B show high
[C/Fe] that is the same feature as the angle-integrated yield. 

The angle-delimited yield strongly depends on which mass elements are
included into the integration. This would be determined
by the abundance mixing in the SN ejecta and by the region where the
next-generation star takes in the metal-enriched gas. To investigate
this issue further, it is required to calculate three-dimensional
evolution of the supernova remnant in the ISM.

(3) {\bf Comparison with the spherical explosion}: 
The angle-integrated yield of the jet-induced explosion is well
reproduced by a spherical SN model applied the mixing-fallback model. This
confirms that the mixing-fallback model approximates the
jet-induced explosion well and that the mixing and fallback in
hypernovae assumed in the mixing-fallback model are actually
achieved in aspherical explosions. 
The abundance ratios between elements synthesized
in different regions (\eg C, O, Mg, and Fe) depend on the hydrodynamical
structure of the explosion, \eg the fallback. Thus, such macroscopic
properties of the jet-induced explosion are represented by the
mixing-fallback model. In particular, $\Min$, $f_{\rm Si}$, and $\Mbh$
in the jet-induced explosion model are represented by $M_{\rm cut}{\rm
(ini)}$, $f$, and $\Mbh$ in the mixing-fallback model, respectively.

On the other hand, the ratios between the explosively-synthesized elements depend on
the thermodynamical properties of the explosion. In particular, [Sc/Fe],
[Ti/Fe], [V/Fe], [Cr/Fe], [Co/Fe] and [Zn/Fe] are enhanced
by the high-entropy environment in the jet-induced explosion, thus 
showing differences from the mixing-fallback model. 
The enhancement of [Sc/Fe] and [Ti/Fe] improve the agreement with the
observations. Note, the enhancement of
[Ti/Fe] relative to the spherical SN model in this paper is larger than
in \cite{mae03b}. This might be because the relativistic jet suppresses
the lateral expansion and enhances the energy concentration.

The enhancement of [Sc/Fe] is also suggested to be obtained by
nucleosynthesis in the $p$-rich ejecta expelled from the innermost
region \citep{pru04,pru05,fro06,iwa06}. However, it seems difficult to
enhance [Zn/Fe] ($\sim0.5$) and [Sc/Fe] ($\sim0$) only with
nucleosynthesis in the $p$-rich ejecta (see \S~5.2 in
\citealt{tom07b}). On the other hand, a jet-induced explosion enhances
[Sc/Fe], [Ti/Fe], and [Zn/Fe] simultaneously and realizes the mixing and
fallback in an energetic explosion. Therefore, I propose that the
jet-induced explosion is a presumable origin of the enhancement of
[Sc/Fe]. The origin would be concluded by future quantitative
investigations including various elements.

Although the models with same $\thj$ are shown in this paper, the
asphericity is likely to be different for each SN. For example, the
observations of supernova remnants show various aspherical structures
(\eg Cassiopeia A, \citealt{hwa04,fes06}; and W49B, \citealt{mic06,keo07}). 
Furthermore, recent observations of the nebular spectra of SNe suggested
that all core-collapse SNe are aspherical explosions
\citep{mae08,mod08}. The degree of asphericity in SNe in the
early universe will be revealed by detailed comparisons of
abundance ratios between the aspherical SN models with different $\thj$ 
and the metal-poor stars which represent the hydrodynamical and/or
thermodynamical properties.

\acknowledgements

This work formed a part of the author's PhD thesis (\citealt{tom07c}, completed in
September 2007).
The author would like to thank H.~Umeda, K.~Maeda, K.~Nomoto, and N.~Iwamoto
for providing the progenitor models and valuable discussion.
Data analysis were in part carried out on the general-purpose PC farm
at Center for Computational Astrophysics, CfCA, of National Astronomical
Observatory of Japan.
The author is supported through the JSPS (Japan Society
for the Promotion of Science) Research Fellowship for Young Scientists.

\appendix

\section{Special relativistic hydrodynamic code}
\label{sec:hydtest}

A two-dimensional special relativistic Eulerian hydrodynamic
code is developed with Marquina's flux formula \citep{don98} and with a
conversion method from the observer frame to the proper frame \citep{mart96}. 
The code applies third order Runge-Kutta method in time of
Shu \& Osher (1988, see also \citealt{alo99}) and second order piecewise
hyperbolic method (PHM) in space of \cite{marq94}.

The equations of special relativistic hydrodynamics are described in
terms of a four-velocity vector field and an energy momentum tensor
\citep{mart94}. Physical quantities in a rest frame (relativistic
rest-mass density: $D$, the $i$-th components of momentum densities: $S_i$,
and energy density: $\tau$) and a comoving frame are related as follows:
\begin{eqnarray}
\nonumber D&=&\rho \Gamma, \\ 
\nonumber S_i&=&\rho h \Gamma^2 v_i, \\ 
\nonumber \tau&=&\rho h \Gamma^2 -p -\rho \Gamma, 
\end{eqnarray}
where the light velocity is set to $c=1$, $\rho$ is the proper rest-mass
density, $v_i$ are the $i$-th components of velocities, 
$\Gamma=1/\sqrt{1-\sum_i v_i^2}$ is the Lorentz factor of the fluid
element with respect to the rest frame, $p$ is the proper pressure, and
$h=1+e_{\rm pr,in}/\rho+p/\rho$ is the specific enthalpy (here, 
$e_{\rm pr,in}$ is the proper internal energy per unit volume).

The basic equations in the special relativistic hydrodynamics are written in
spherical polar coordinates ($r, \theta$) as
\begin{eqnarray}
 {\partial D\over{\partial t}}+{\partial (r^2Dv_r)\over{\partial (r^3/3)}}+{1\over{r}}{\partial (\sin{\theta}Dv_\theta) \over{\partial (-\cos{\theta})}}&=&0, \\ 
  {\partial S_r\over{\partial t}}+{\partial
 (S_rv_r+p)\over{\partial r}}+{1\over{r}}{\partial
 (\sin{\theta}S_rv_\theta)\over{\partial (-\cos{\theta})}}
&=&-2{S_rv_r\over{r}}+{S_\theta v_\theta\over{r}}+g_r \rho, \\
  {\partial S_\theta \over{\partial t}}+{\partial (r^2S_\theta
 v_r)\over{\partial (r^3/3)}}+{1\over{r}}{\partial (S_\theta v_\theta
 +p) \over{\partial \theta}}
&=&-{1\over{r}}{\cos{\theta}\over{\sin{\theta}}}-{S_\theta v_r\over{r}}+g_\theta \rho, \\
 {\partial \tau\over{\partial t}}+{\partial
  \{r^2(S_r-Dv_r)\}\over{\partial(r^3/3)}}+{1\over{r}}{\partial
  \{\sin{\theta}(S_\theta -Dv_\theta)\}\over{\partial
  (-\cos{\theta})}}
&=&(g_r v_r+g_\theta v_\theta)\rho,
\end{eqnarray}
where $g_i$ ($i=r, \theta$) are the $i$-th gravitational acceleration
components. 
The equations are an equation of continuity, momentum
conservation equations, an energy conservation equation.
The gravitational potential includes the contributions of the
self-gravity and the central remnant. I test two methods to include the
self-gravity as follows: (1) the Poisson equation is approximated with
an integration of spherical harmonics \citep{hac86} and (2) the sparse
banded matrix resulting from differencing the Poisson equation is solved
with a bi-conjugate gradient stabilized (BiCGSTAB) method
\citep{bar94}\footnote{Subroutines are available at Netlib
(\url{http://www.netlib.org/}).} with modified incomplete LU (MILU)
factorization. Since the results are consistent, I
apply the former method in the calculations.

\subsection{One-dimensional Riemann problems}
\label{sec:1Dshocktest}

The code is tested with one-dimensional shock tube problems (\eg
\citealt{mart94}). I set the computational region 
$0\leq x \leq 1$ where $c=1$ that is divided
into two regions, \ie left and right regions, at $x=0.5$. The density
($\rho$), pressure ($p$), and velocity ($v$) of left and right regions are
represented by L and R subscripts, respectively.
The time variation is derived by the analytical solution of the Riemann
problem \citep{mart94}. I test three problems with $10^3$ uniform meshes
($\Delta x =1/10^3$) and
a constant adiabatic index. The initial states are as follows:

Problem (a) $\gamma=5/3$
\begin{eqnarray}
\nonumber (\rho_{\rm L},p_{\rm L},v_{\rm L})&=&(1, 10^3, 0) \ {\rm at} \ 0.0\leq x \leq 0.5, \\
\nonumber (\rho_{\rm R},p_{\rm R},v_{\rm R})&=&(1, 0.01, 0) \ {\rm at} \ 0.5\leq x \leq 1.0.
\end{eqnarray}

Problem (b) $\gamma=4/3$
\begin{eqnarray}
\nonumber (\rho_{\rm L},p_{\rm L},{\Gamma}_{\rm L})&=&(1, 1, 10^3) \ {\rm at} \ 0.0\leq x \leq 0.5, \\
\nonumber (\rho_{\rm R},p_{\rm R},{\Gamma}_{\rm R})&=&(1, 10^{6}, 1) \ {\rm at} \ 0.5\leq x \leq 1.0. 
\end{eqnarray}

Problem (c) $\gamma=4/3$
\begin{eqnarray}
\nonumber (\rho_{\rm L},p_{\rm L},v_{\rm L})&=&(1, 10, -0.9) \ {\rm at} \ 0.0\leq x \leq 0.5, \\
\nonumber (\rho_{\rm R},p_{\rm R},v_{\rm R})&=&(10, 100, 0.9) \ {\rm at} \ 0.5\leq x \leq 1.0.
\end{eqnarray}

\begin{figure}[h]
\epsscale{1.}
	 \plotone{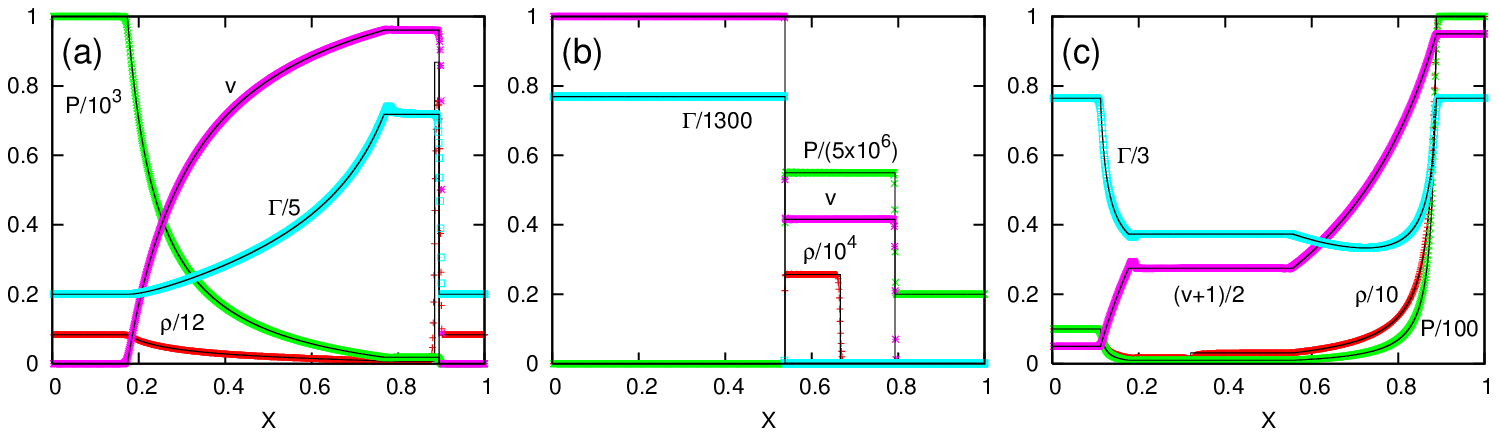}
\caption{Comparison between the analytical solutions of the Riemann
	 problems ({\it black solid line}) and the results of the
	 special relativistic hydrodynamics calculations (density: {\it
	 red marks}, pressure: {\it green marks}, velocity: {\it magenta
	 marks}, Lorentz factor: {\it cyan marks}). \label{fig:2Drel1Dshock}}
\end{figure}

The analytical solutions are reproduced by the calculations
 (Figs.~\ref{fig:2Drel1Dshock}abc).
The agreements confirm that the code correctly solves
left- and right-oriented rarefaction and shock waves.

\subsection{Two-dimensional shock tube problem}

\begin{figure}[h]
\epsscale{.55}
  \plotone{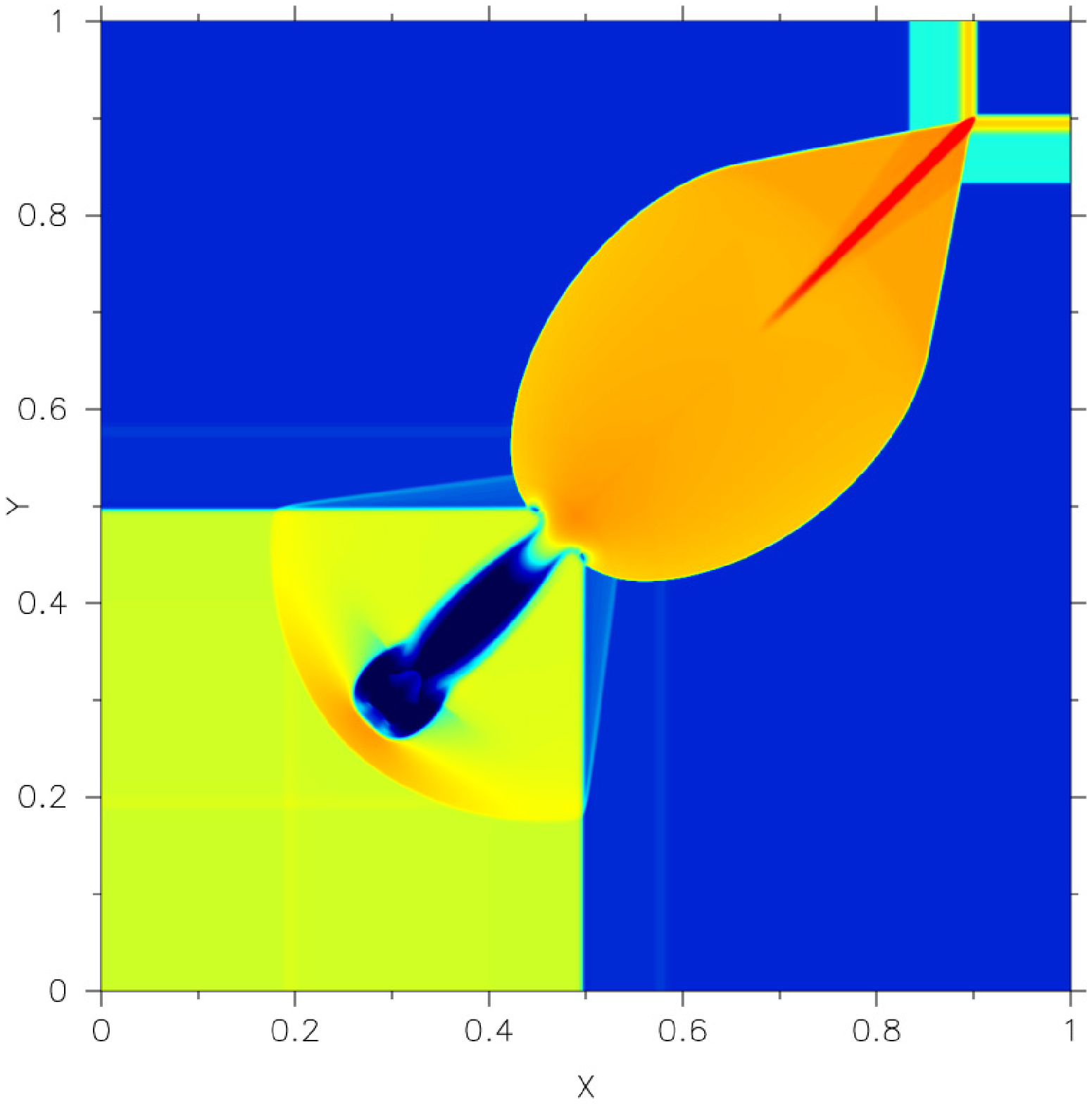}
\caption{Two-dimensional shock tube problem at $t=0.4$ with
 $10^3\times10^3$ square uniform meshes. Color contours of the logarithm
 of the rest mass density are plotted. \label{fig:2Drel2Dshock}}
\end{figure}

The two-dimensional shock tube problem was proposed by \cite{del02} and
repeated by subsequent studies (\eg \citealt{luc04,zha06,miz06}). I test the problem
with $10^3\times10^3$ square uniform meshes. In this problem, the
two-dimensional region ($0\leq x \leq 1, 0 \leq y \leq 1$) are
divided into the following four regions: 
\begin{eqnarray}
\nonumber (\rho,p,v_x,v_y)&=&(0.50,1.00,0.00,0.00) \ {\rm at} \ 0.0\leq x \leq 0.5, 0.0\leq y \leq 0.5, \\
\nonumber (\rho,p,v_x,v_y)&=&(0.10,1.00,0.99,0.00) \ {\rm at} \ 0.0\leq x \leq 0.5, 0.5\leq y \leq 1.0, \\
\nonumber (\rho,p,v_x,v_y)&=&(0.10,0.01,0.00,0.00) \ {\rm at} \ 0.5\leq x \leq 1.0, 0.0\leq y \leq 0.5, \\
\nonumber (\rho,p,v_x,v_y)&=&(0.10,1.00,0.00,0.99) \ {\rm at} \ 0.5\leq x \leq 1.0, 0.5\leq y \leq 1.0, 
\end{eqnarray}
where $v_i$ ($i=x,~y$) are the $i$-th components of velocities and the adiabatic index is
$\gamma=5/3$. Figure~\ref{fig:2Drel2Dshock} shows color contours of $\log(\rho)$ at $t=0.4$. 

\subsection{Double Mach reflection problem}

\begin{figure}[h]
\epsscale{.75}
	 \plotone{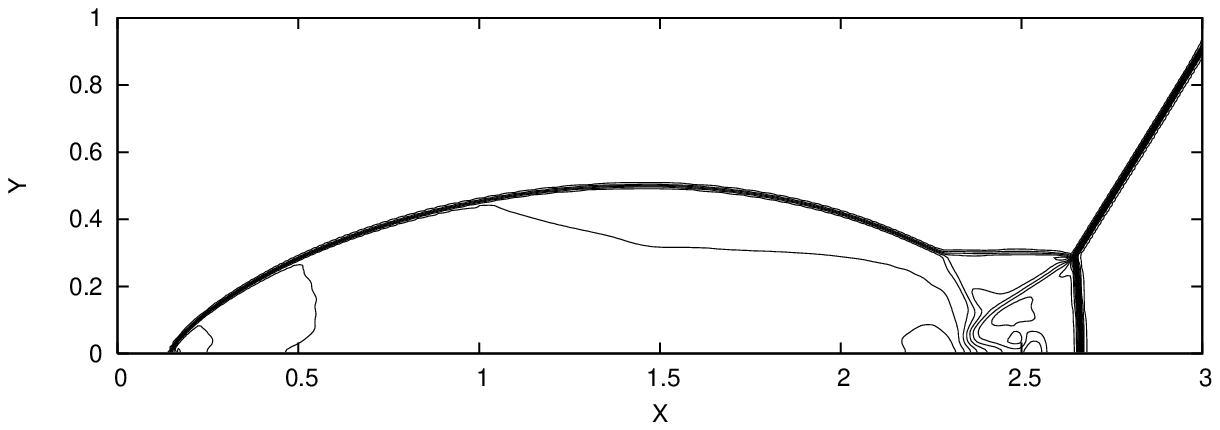}
\caption{Double-Mach reflection problem at $t=4.0$ with
 $512\times128$ uniform square meshes. Thirty equally-spaced contours of
	 the logarithm of the rest
	 mass density are plotted.\label{fig:2Drel2Dmach}}
\end{figure}

\cite{woo84} introduced a double Mach reflection problem of a strong shock in
the Newtonian case and the problem was extended to the special relativistic case 
\citep{zha06}. I test the code with the same
initial conditions as in the previous studies \citep{zha06,miz06}. The
computational region is a $0\leq x \leq 4, 0 \leq y \leq 1$
rectangle captured by $512\times128$ uniform square meshes. 
The density and pressure of the unshocked gas, the adiabatic index,
and the classical Mach number ($M_{\rm c}\equiv V_{\rm S}/c_{\rm s}$, where
$V_{\rm S}$ is a shock velocity and $c_{\rm s}$ is a sound speed) are
set to be $\rho=1.4$, $p=0.0025$, $\gamma=1.4$, and $M_{\rm c}=10$
respectively. Thus, the shock velocity is $V_{\rm S}=0.4984$ and the
density, pressure, and velocity of the post-shock gas are $\rho=8.564$,
$p=0.3804$, and $v=0.4247$. The shock front
is initially set to cross the $x$ axis at $x=1/6$ with an angle of
$60^\circ$. The initial conditions are as follows:
\begin{eqnarray}
\nonumber (\rho,p,v_x,v_y)&=&(1.4,0.0025,0.00,0.00) \ {\rm at} \ x - y \tan{60^\circ} < 1/6, \\
\nonumber (\rho,p,v_x,v_y)&=&(8.564,0.3804,0.4247 \sin{60^\circ}, 0.4247 \cos{60^\circ}) \ {\rm at}  \ x - y \tan{60^\circ} > 1/6.
\end{eqnarray}
The boundary conditions are inflow of the post-shock gas at
$x=0$, at $0\leq x \leq 1/6$ on $y=0$ axis, and at 
$0\leq x \leq (1/6+\tan{60^\circ}+V_{\rm S}t/\sin{60^\circ})$ on $y=1$ axis, 
reflective boundaries at $1/6<x<4$ on $y=0$ axis, and freely inflow and
outflow conditions at the other boundaries. Figure~\ref{fig:2Drel2Dmach}
shows thirty-level iso-surfaces of $\log(\rho)$ at $t=4.0$. 

\subsection{Emery step problem}

\begin{figure}[h]
\epsscale{.75}
	 \plotone{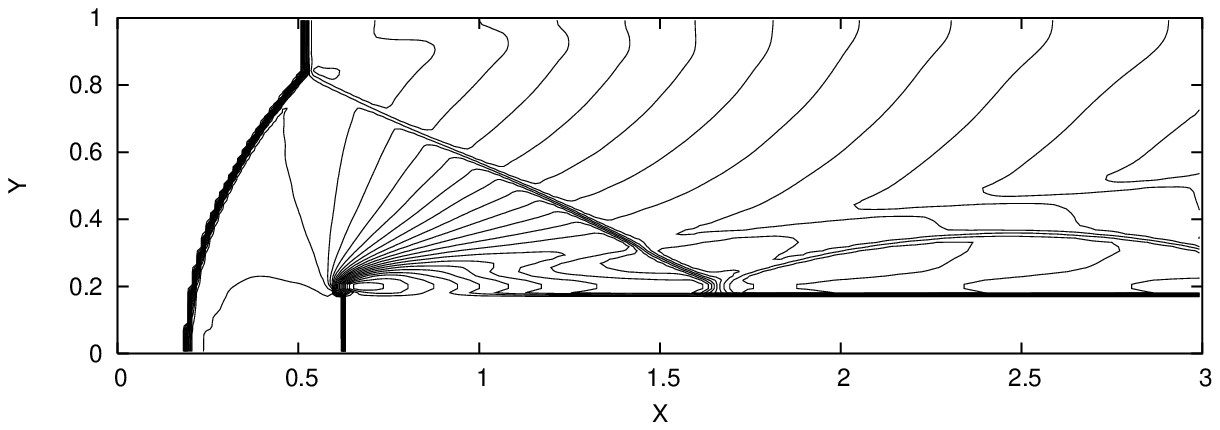}
\caption{Emery step problem at $t=4.0$ with $240\times80$ uniform square
	 meshes. Thirty equally-spaced contours of the logarithm of the rest
	 mass density are plotted. \label{fig:2Drel2Demery}}
\end{figure}

A wind tunnel problem with a reflecting step, called the Emery step problem,
was introduced by \cite{eme68} and repeated by subsequent studies
\citep{zha06,miz06}. The computational domain is a 
$0\leq x \leq 3, 0 \leq y \leq 1$ rectangle captured by $240\times80$
uniform square meshes. The domain is filled by a gas with $\rho=1.4$,
$p=0.1534$, and $v=0.999$. The pressure is set so that $M_{\rm c}=3$.
The wind tunnel has the reflecting step locating at $0.6\leq x$ and 
$y\leq 0.2$. The following boundary conditions are applied: the upper
and lower boundaries are reflective and the boundaries at $x=0$ and
$x=3$ are open. Figure~\ref{fig:2Drel2Demery} shows thirty-level
iso-surfaces of $\log(\rho)$ at $t=4.0$. 

\section{Nucleosynthesis calculation}
\label{sec:nuctest}

The two-dimensional special relativistic hydrodynamical calculation
does not include nuclear energy releases and applies a constant
adiabatic index $\gamma=4/3$. The proper internal energy is written as
$e_{\rm pr,in}=p/(\gamma-1)$. Temperature $T$ is derived with an analytical equation
of state including radiation and ${\rm e}^-$-${\rm e}^+$ pair (\eg
\citealt{fre99}) as follows:
\begin{equation}
 e_{\rm pr,in}=aT^4 \left\{1+{7\over{4}}\cdot{T_9^2\over{T_9^2+5.3}}\right\}.
\end{equation} 
where $a=7.57\times10^{-15}$ ${\rm ergs~cm^{-3}~K^{-4}}$ is the
radiation-density constant and $T_9=T/10^9$ K.

\begin{figure}[h]
\epsscale{.47}
  \plotone{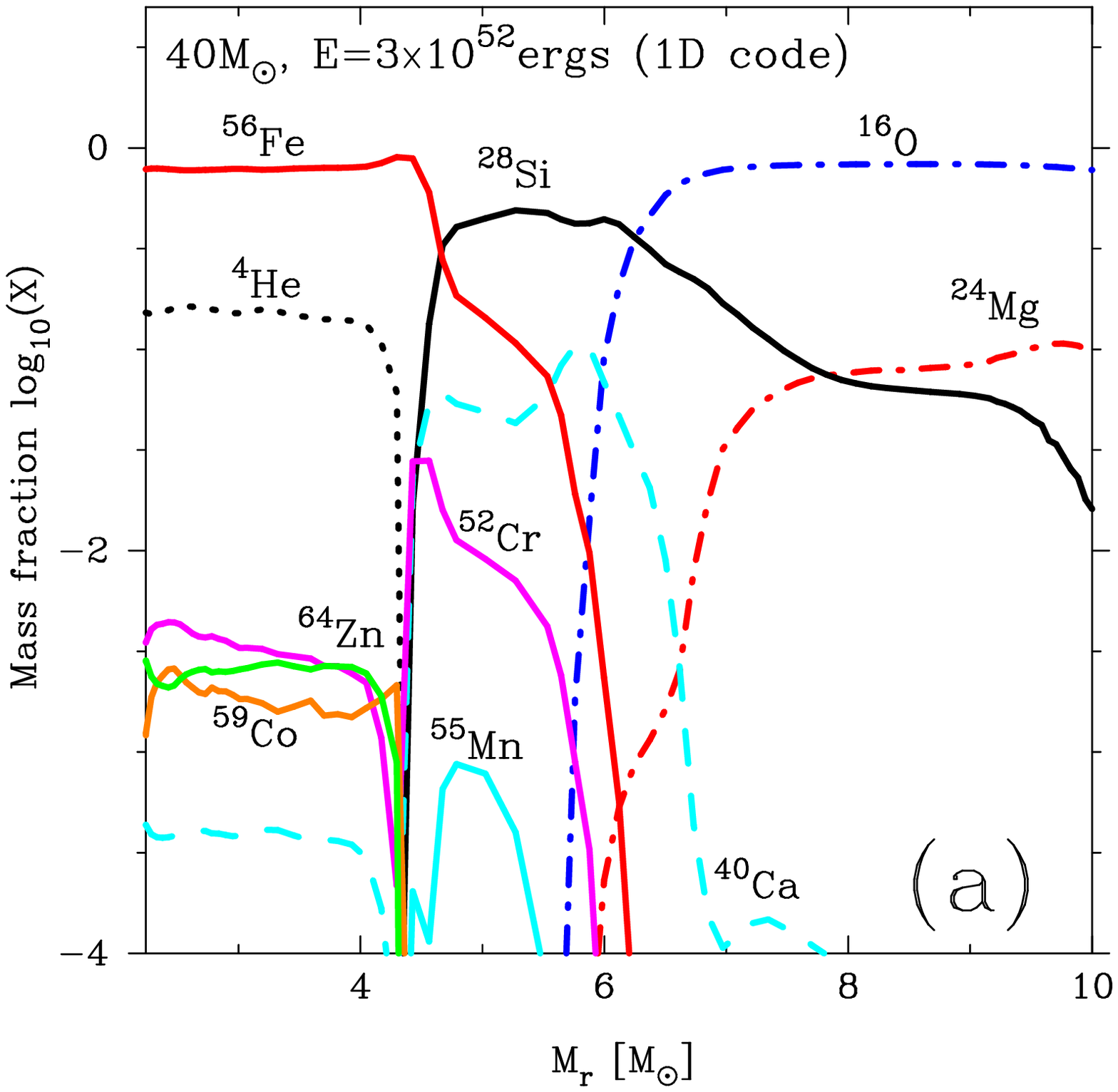}
  \plotone{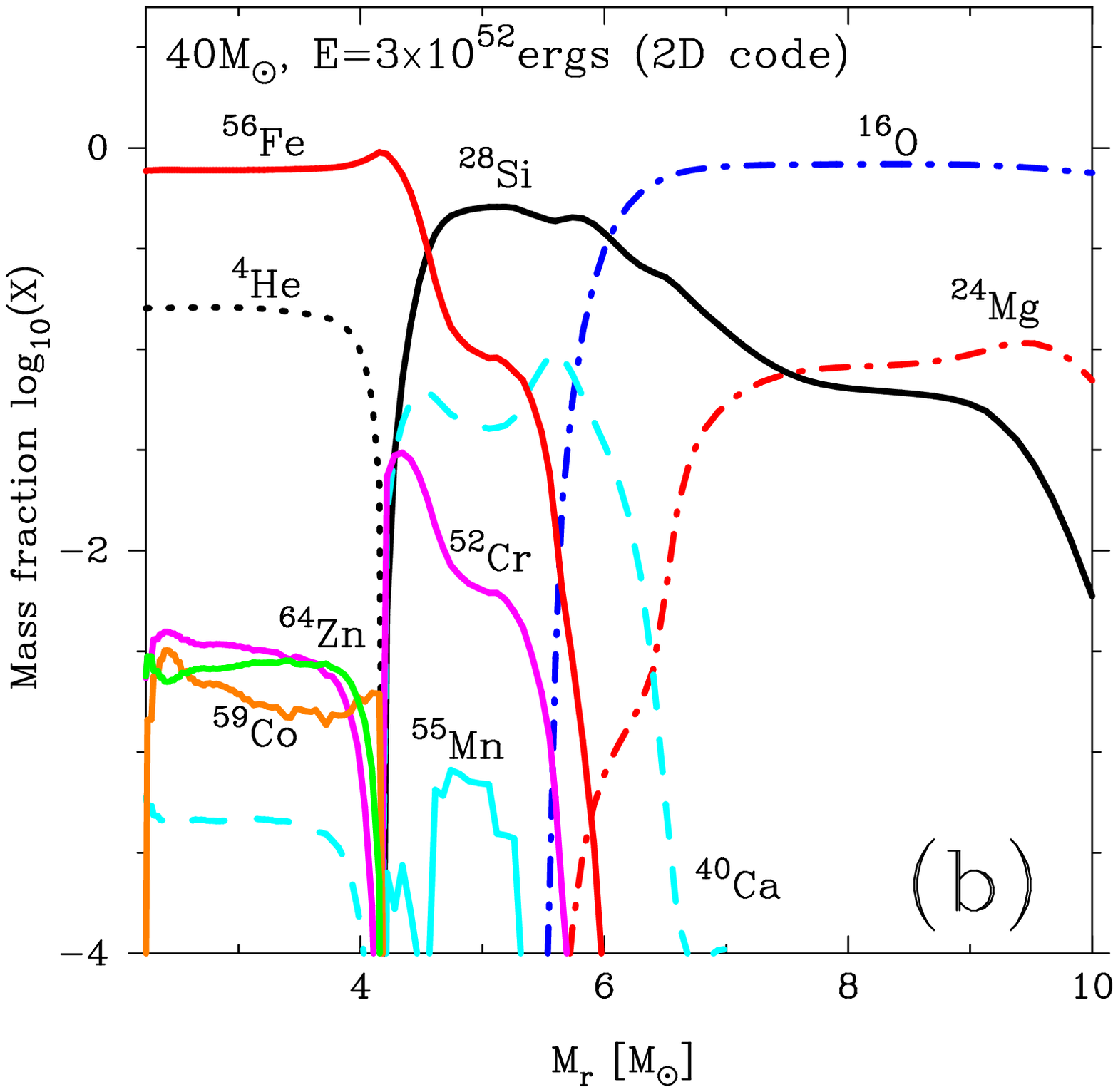}
\caption{Abundance distributions after the non-relativistic spherical
  explosion of a $40\Msun$ star with $3\times10^{52}$ ergs. The
 post-processing calculations are performed with the thermodynamic
 histories obtained by (a) the one-dimensional spherical hydrodynamics
  code and (b) the two-dimensional special relativistic
 hydrodynamical code. \label{fig:2Drel2Dnuc}}
\end{figure}

I calculate explosive nucleosynthesis in a non-relativistic spherical explosion of a $40\Msun$
star with $3\times10^{52}$~ergs. The post-processing calculations are
performed with the thermodynamic histories obtained by the two-dimensional special
relativistic Eulerian hydrodynamics code and a one-dimensional Lagrangian hydrodynamics code
used in \cite{ume02,ume05} and \cite{tom07b}.
The one-dimensional Lagrangian hydrodynamical
calculation includes nuclear energy releases from the $\alpha$-network
and the equation of state includes the gas, radiation,
${\rm e}^-$-${\rm e}^+$ pair \citep{sug75}, Coulomb interactions between ions and electrons, and
phase transition \citep{nom82,nom88}.
The abundance distributions derived by both calculations are consistent
(Figs.~\ref{fig:2Drel2Dnuc}ab). This consistency justifies the results
of the two-dimensional special relativistic hydrodynamics and
nucleosynthesis calculations even with the simple assumptions.

\section{Jet injection}
\label{sec:jetin}

The jet characterized by the five parameters, $\Ed$, $\Et$, $\thj$,
$\Gj$, and $\fth$, is injected from the inner boundary at $\Rin$.
The jet injection is performed by putting the velocities 
($v_{r,{\rm jet}}$, $v_{\theta,{\rm jet}}$), pressure ($p_{\rm jet}$),
and density ($\rho_{\rm jet}$) of the jet at the inner boundary as the
boundary condition. $v_{r,{\rm jet}}$, $v_{\theta,{\rm jet}}$, 
$p_{\rm jet}$, and $\rho_{\rm jet}$ are derived from the five parameters.

The superficial area of the jet injection is 
$A_{\rm jet}=4\pi\Rin^2(1-\cos\thj)$ and thus the energy deposition rate
is $\Ed=A_{\rm jet}f_{e,{\rm jet}}$, where $f_{e,{\rm jet}}$ is an energy
density flux of the jet. $f_{e,{\rm jet}}$ is written with
the jet properties as 
$f_{e,{\rm jet}}=S_{r,{\rm jet}}-D_{\rm jet}v_{r,{\rm jet}}$, where 
$S_{r,{\rm jet}}=\rho_{\rm jet} h_{\rm jet} \Gj^2 v_{r,{\rm jet}}$,
$D_{\rm jet}=\rho_{\rm jet} \Gj$, and $h_{\rm jet}$ is the specific
enthalpy of the jet. Although the energy density flux at the inner
boundary ($f_e$) depends on the properties of the jet and the infalling
matter and changes with time, $f_e$ is
equivalent to $f_{e,{\rm jet}}$ when the jet is injected freely.

On the other hand, the energy density per unit volume is 
$e_{\rm jet}=\tau_{\rm jet}=\rho_{\rm jet} h_{\rm jet} \Gj^2 -p_{\rm jet}-\rho_{\rm jet} \Gj$. 
The energy density
consists of the thermal energy density ($e_{\rm th,jet}$) and the
kinetic energy density ($e_{\rm kin,jet}$). Applying 
$h_{\rm jet}=1+e_{\rm pr,in,jet}/\rho_{\rm jet}+p_{\rm jet}/\rho_{\rm jet}$
and $e_{\rm pr,in,jet}=p_{\rm jet}/(\gamma-1)$, $e_{\rm th,jet}$ and 
$e_{\rm kin,jet}$ are written by the density $\rho_{\rm jet}$, pressure
$p_{\rm jet}$, Lorentz factor $\Gj$, and adiabatic index $\gamma$ of the
jet as follows:
\begin{eqnarray}
 e_{\rm kin,jet}=(1-\fth) e_{\rm jet}=\rho_{\rm jet} \left(\Gj^2-\Gj\right), \\
 e_{\rm th,jet}=\fth e_{\rm jet}=p_{\rm jet} \left({\gamma\over{\gamma-1}}\Gj^2-1\right).
\end{eqnarray}

Since $e_{\rm jet}=f_{e,{\rm jet}}/v_{r,{\rm jet}}-p_{\rm jet}$, the
velocities, pressure, and density of the jet are written as follows:
\begin{eqnarray}
 v_r&=&\sqrt{1-1/\Gj^2},\\
 v_\theta&=&0,\\
 p_{\rm jet}&=&{{\fth \Ed}\over{v_r A_{\rm jet}\left({\gamma\over{\gamma-1}}\Gj^2-1+\fth\right)}},\\
 \rho_{\rm jet}&=&{(1-\fth)\left({\Ed \over{ v_r  A_{\rm jet}}}-p_{\rm jet}\right)\over{\Gj^2-\Gj}}.
\end{eqnarray}

\end{document}